\documentclass[aps,prx,superscriptaddress]{revtex4-2}
\usepackage{graphicx}
\usepackage{xcolor}
\usepackage{soul}
\usepackage[normalem]{ulem} 
\usepackage{array}
\usepackage{amsmath}
\usepackage{lineno}

\begin{document}


\title{Gated InAs quantum dots embedded in surface acoustic wave cavities \\for low-noise optomechanics}


\author{Zixuan Wang}
\email{zixuan.wang@colorado.edu}
\affiliation{National Institute of Standards and Technology, Boulder, Colorado 80305, USA}
\affiliation{Department of Physics, University of Colorado, Boulder, CO 80309, USA}
\author{Ryan A. DeCrescent}
\affiliation{National Institute of Standards and Technology, Boulder, Colorado 80305, USA}
\author{Poolad Imany}
\affiliation{National Institute of Standards and Technology, Boulder, Colorado 80305, USA}
\affiliation{Department of Physics, University of Colorado, Boulder, CO 80309, USA}
\affiliation{Icarus Quantum Inc., Boulder, Colorado 80302, USA}
\author{Joey T. Bush}
\affiliation{National Institute of Standards and Technology, Boulder, Colorado 80305, USA}
\affiliation{Department of Physics, University of Colorado, Boulder, CO 80309, USA}
\author{Dileep V. Reddy}
\affiliation{National Institute of Standards and Technology, Boulder, Colorado 80305, USA}
\affiliation{Department of Physics, University of Colorado, Boulder, CO 80309, USA}
\author{Sae Woo Nam}
\affiliation{National Institute of Standards and Technology, Boulder, Colorado 80305, USA}
\author{Richard P. Mirin}
\affiliation{National Institute of Standards and Technology, Boulder, Colorado 80305, USA}
\author{Kevin L. Silverman}
\email{kevin.silverman@nist.gov}
\affiliation{National Institute of Standards and Technology, Boulder, Colorado 80305, USA}

\date{\today}

\begin{abstract}
Self-assembled InAs quantum dots (QDs) are promising optomechanical elements due to their excellent photonic properties and sensitivity to local strain fields. Microwave-frequency modulation of photons scattered from these efficient quantum emitters has been recently demonstrated using surface acoustic wave (SAW) cavities. However, for optimal performance, a gate structure is required to deterministically control the charge state and reduce charge noise of the QDs. Here, we integrate gated QDs and SAW cavities using molecular beam epitaxy and nanofabrication. We demonstrate that with careful design of the substrate layer structure, integration of the two systems can be accomplished while retaining the optimal performance of each subsystem. These results mark a critical step toward efficient and low-noise optomechanical systems that truly leveraging the excellent properties of semiconductor QDs.
\end{abstract}


\maketitle

\section*{Introduction}

Microwave-to-optical quantum transducers are crucial to unlock the potential of future quantum networks, leveraging the unique strengths of diverse quantum systems \cite{Han2021,Mirhosseini2020}. They bridge the gap between microwave frequencies that are characteristic of superconducting \cite{Arute2019,Kim2023} or spin-based \cite{Eriksson2004,Kloeffel2013} quantum processors and optical frequencies that are ideal for transmitting quantum information over long distances \cite{VanLeent2020,Liao2017}. These devices would thus allow the exchange of quantum information between quantum nodes over distances that would otherwise be infeasible. An ideal transducer would convert microwave signals to optical frequencies with 100\% efficiency and no added noise. However, a five-orders-of-magnitude energy gap between microwave and optical domains makes it challenging to build such transducers with high efficiency. Heralding protocols can largely bypass this obstacle, improving fidelity at the expense of communication rate. Therefore, minimizing added noise is perhaps a more important immediate consideration for such transducers \cite{Jiang2023}. 

Among various approaches for quantum transduction, hybrid systems involving mechanical resonators (confined phonons) are promising \cite{Higginbotham2018}. These electro-optomechanical systems take advantage of the slower speed of phonons compared to photons, resulting in phonons at gigahertz (GHz) frequencies with wavelengths comparable to optical photons. This is key for mediating interactions across the energy gap. It has been shown that phonons interact well with various types of quantum systems \cite{Delsing2019}. For example, they piezoelectrically interact with superconducting qubits \cite{Chou2020} and parametrically modulate various optical systems such as optical cavities \cite{Brubaker2022} and single photon emitters (SPEs) in 2D materials \cite{Iikawa2019,lazic2019,Patel2022}, defect centers \cite{Golter2016,Hernandez-Minguez2021,Whiteley2019}, and semiconductor quantum dots (QDs) \cite{Pustiowski2015,Imany2022,Metcalfe2010,Weib2021,Spinnler2023b}. However, integrating two standalone electromechanical and optomechanical systems is challenging due to the added system complexity and possible incompatibility \cite{Clerk2020}.    

In this paper, we address this integration challenge in an electro-optomechanical system involving gated InAs QDs and SAW resonator phonons. Our system combines excellent optical performance in gated InAs QDs with state-of-the-art surface acoustic wave (SAW) resonators on GaAs, fully incorporating microwave and DC electronics, optics, and mechanics in a single platform. In general, this development is crucial to fully realize the advantages of QDs as optical two-level systems for optomechanics, including single-phonon generation \cite{Sollner2016} and acousto-optic preparation of exciton or biexciton states \cite{Kuniej2024} which have so far been only theoretically proposed. In addition, interactions between acoustics and other types of SPEs (e.g., diamond vacancies) have recently caught attention as useful and interesting systems \cite{McCullian2024,Lukin2020,Kepesidis2013}. This work brings semiconductor QDs, with their many desirable properties, up to par with these other systems. When combined with superconducting qubits \cite{Bienfait2019,aref2016}, these results pave the way for low-noise and efficient conversion of quantum information between microwave and optical domain.

Self-assembled InAs QDs have emerged as highly favorable SPEs and offer many benefits over other systems, such as trapped ions, color centers, vacancies, or defects. For example, previous work has demonstrated state-of-the-art end-to-end photon collection efficiencies of 57\% \cite{Tomm2021}, two-photon interference visibilities of about 99\% \cite{Zhai2022}, and repetition rates of approximately 1 GHz \cite{Tomm2021}. InAs QDs are embedded in semiconductor materials and are compatible with various fabricated photonic structures for efficient photon extraction \cite{Hughes2005,Luxmoore2013,Reithmaier2004,Strauf2006}. Their electrostatic environment and charge state can be controlled by simple gate structures \cite{Warburton2013}. With deterministic trapping of a single charge or hole, InAs QDs enable spin-photon entanglement \cite{Sun2014} and remote spin-spin entanglement \cite{Delteil2016}, which are crucial for quantum information processing applications. Finally, they exhibit a relatively simple energy level structure, approximating a two-level system, allowing for deterministic excitation with very simple optical pumping schemes. This provides a promising route to generate single phonons that are important for phononic quantum technologies \cite{Sollner2016}. In contrast, atoms and ions require active trapping mechanisms and complex optical pumping schemes for cooling and state preparation \cite{Slodička2013}, and suffer from poor photon collection efficiencies into fiber \cite{Leent2022}. Color centers, vacancies and defects are other solid-state SPEs with promising properties, but they suffer from non-unity quantum efficiencies and often fast phonon-assisted emission pathways \cite{Bradac2019}.

InAs QDs are also promising choices for optomechanical devices as they are very sensitive to local strain fields \cite{Sun2013} due to the large deformation potential of GaAs ($\sim$$10^{14}-10^{15}$ Hz) \cite{Metcalfe2010}. This sensitivity is exemplified by large optomechanical single-phonon coupling rates on the order of a few MHz \cite{DeCrescent2022,Spinnler2023a}, exceeding state-of-the-art systems based on optomechanical crystals \cite{Forsch2020,Han2021}. Moreover, with GHz-frequency phonon modes, mechanical motion can be passively cooled to the ground state while the QD scatters photons at GHz rates under extremely low optical pump powers ($\sim$100 pW) \cite{Najer2019}, suggesting the possibility for low-noise operation \cite{Spinnler2023a}. Although previous results are promising, the lack of gated QDs limited optomechanical performance in two ways: 1) reduced interaction duty cycle between the QD and the resonant pump laser as the QD randomly hops between different charge states; 2) inhomogeneous broadening of the QD linewidth due to moving charges nearby. Both detrimental impacts lead to lower optomechanical cooperativity as well as transduction efficiency. These gate structures often involve a thin n-doped GaAs layer (GaAs:Si) close below the QD layer and either a layer of metal (Schottky structure) or p-doped GaAs (p-i-n structure) above \cite{Warburton2013}. By applying an electric potential across the gate, the net charge of the QD is controlled via the Coulomb blockade, and the charge noise is significantly reduced as excess charges get depleted. The importance of integrating charge controlled QDs with the state-of-the-art resonators cannot be overstated, with evidence in the years of researches that led to QD devices surpassing the performance of down-conversion sources for pure and indistinguishable single-photon generation \cite{Najer2019,Somaschi2016}. The main challenge in integrating SAW resonators with gated QDs is the impact of the doped semiconductors on the creation and propagation of SAWs. 

\begin{figure}[h!]
\centering\includegraphics[width=0.9\linewidth]{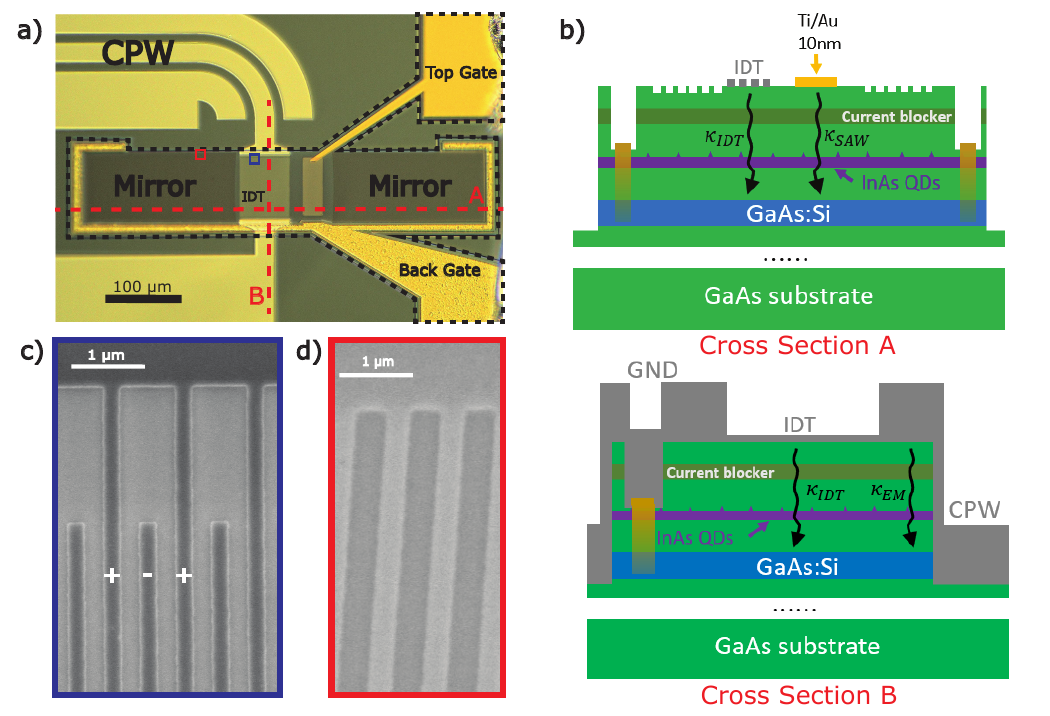}
\caption{Illustration and micrographs of the integrated device. a) Optical micrograph of the fabricated device integrating gated QDs with SAWs. CPW: Co-planar waveguide. IDT: Interdigitated transducers. b) Illustration of the cross-sections A and B of the proposed design, corresponding to the horizontal and vertical red dashed lines in panel a. The main integration challenge is the loss introduced to the launching via the IDT ($\kappa_{IDT}$) and propagation of the SAW due to the doped GaAs layer below surface ($\kappa_{SAW}$). The microwave transmission through the coplanar waveguide will also be affected, introducing an effective loss $\kappa_{EM}$. As shown in b), the doped layer is etched away outside the black dashed region indicated in panel a) to reduce extra loss introduced to the microwave transmission line. A thin film of Ti/Au is evaporated inside the SAW resonator forming the top gate for the Schottky structure. An ohmic contact is made to the n-doped GaAs layer by Ni and AuGe, serving as the back gate. c) Scanning electron microscope (SEM) image of the Al IDT fingers. d) SEM image of the etched SAW DBR mirror. A slight curvature is added to reduce phonon diffraction loss.}
\label{fig1}
\end{figure}

\section*{Results}
\subsection*{Integration Challenges}

Our integrated device is shown in Fig 1a), with cross-section schematics in Fig 1b) and plan-view scanning-electron micrographs (SEMs) in Figs. 1c) and 1d). The SAWs are launched by metallic interdigitated transducers (IDTs; Fig. 1c) and then confined within etched Bragg reflectors (`mirrors'; Fig. 1d) that define our SAW cavity. The introduction of a buried doped layer, required for QD gating, can adversely affect the system in several ways. We specifically address three main concerns (illustrated in Fig. 1b): 1) SAW propagation losses $\kappa_{SAW}$; 2) piezoelectric coupling between the IDT  and SAWs $\kappa_{IDT}$; 3) and microwave transmission between the microwave source and the device though the coplanar waveguide (CPW) $\kappa_{EM}$. We first address the SAW attenuation, $\kappa_{SAW}$. The effect of a thin doped semiconductor on SAW propagation has been extensively studied in the context of two-dimensional electron gasses \cite{Simon1996,Wixforth1986}: a doped  semiconductor will introduce attenuation $\kappa$ and velocity shift $\Delta V_s$ to the SAW field that satisfy the relations
\begin{equation}
\Delta V_s = \frac{\alpha^2/2}{1+\sigma^2_{xx}/\sigma^2_m},\quad \kappa = \frac{\alpha^2q\sigma_{xx}/2\sigma_m}{1+\sigma^2_{xx}/\sigma^2_m}
\label{kappa}
\end{equation}
where $\sigma_{xx}$ is the conductivity of the doped layer, $\alpha^2$ and $\sigma_m$ are system-specific coefficients that depend on the depth of the doped layer, and $q$ is the wavevector of the SAW. Velocity shifts are easily compensated by changing the SAW resonator length. The attenuation, however, reduces the phonon lifetime and the resonator’s quality factor. As shown in Eqn. \ref{kappa}, the attenuation strongly depends on the conductivity $\sigma_{xx}$ and thus the doping concentration. Negligible SAW attenuation can be achieved when $\sigma_{xx}\gg\sigma_m$, which can be done with a high doping concentration. Specifically, here we designed the doped layer (GaAs:Si) with a doping concentration of $2\times10^{18} \ \text{cm}^{-3}$ and a thickness of 47 nm. Hall measurements reveal a conductivity $\sigma_{xx}$ of $10^5$ S/m, and we estimate the value of $\sigma_{m}$ to be on the order of 10 S/m based on literature \cite{Wixforth1986}. Therefore, we are indeed operating in the low SAW loss regime. Numerical and analytical calculation further shows that these conductivity values corresponding to acoustic loss rate on the order of 100 Hz (Supplementary Information). In comparison, current state-of-the-art GHz SAW resonators have loss rates on the order of 100 kHz due to mostly bulk scattering losses. Therefore, the acoustic attenuation introduced by the doped layer can be negligible with a sufficiently high doping concentration.

We next address the effect of the n-doped layer on SAW launching efficiency, $\kappa_{IDT}$. In general, the n-doped layer will suppress the launching of SAWs by screening the microwave field applied to the IDT \cite{Yuan2017,Yuan2022}. The impact will decrease as the n-doped layer is deeper in the substrate. However, there is a limit on the depth of the n-doped layer: it needs to stay close to the QDs for optimum depletion of excess charges, and the QDs need to be located near the GaAs surface where the strain field is larger. We numerically evaluate the impact on SAW launching efficiency with the n-doped layer located at different depths (Supplementary Information). The conductivity and thickness of the layer are fixed to experimentally derived values. The coupling constant $k^2$ begins to drop dramatically when the n-doped layer is within approximately 400 nm of the IDT. Considering constraints on the QD depth, we choose a n-doped layer depth of 360 nm, which provides decent QD-SAW coupling while retaining 80\% of the coupling constant compared to that of bulk semi-insulating GaAs substrates (Supplementary Information).     

Finally, the n-doped layer will also affect the microwave power delivered to the IDT via the co-planar waveguide (CPW), $\kappa_{EM}$. Due to the proximity of the n-doped layer to the surface (several hundred nanometers), the microwave mode in the CPW (with dimensions on the order of a few micrometers) is modified, thus reducing the characteristic transmission line impedance from approximately 50 $\Omega$ to 3 $\Omega$. The deviation of the impedance will reflect approximately 88\% of the microwave signals before reaching the device, and further reduce the overall SAW launching efficiency. The solution for this is simple: the n-doped layer is unnecessary below the CPW structure and can be removed under the microwave circuits to recover the transmission line impedance. As shown in Fig 1b), the buried n-doped layer exists only in the region within the black dotted line, and is etched away everywhere else. 

\begin{figure}[t!]
\centering\includegraphics[width=0.9\linewidth]{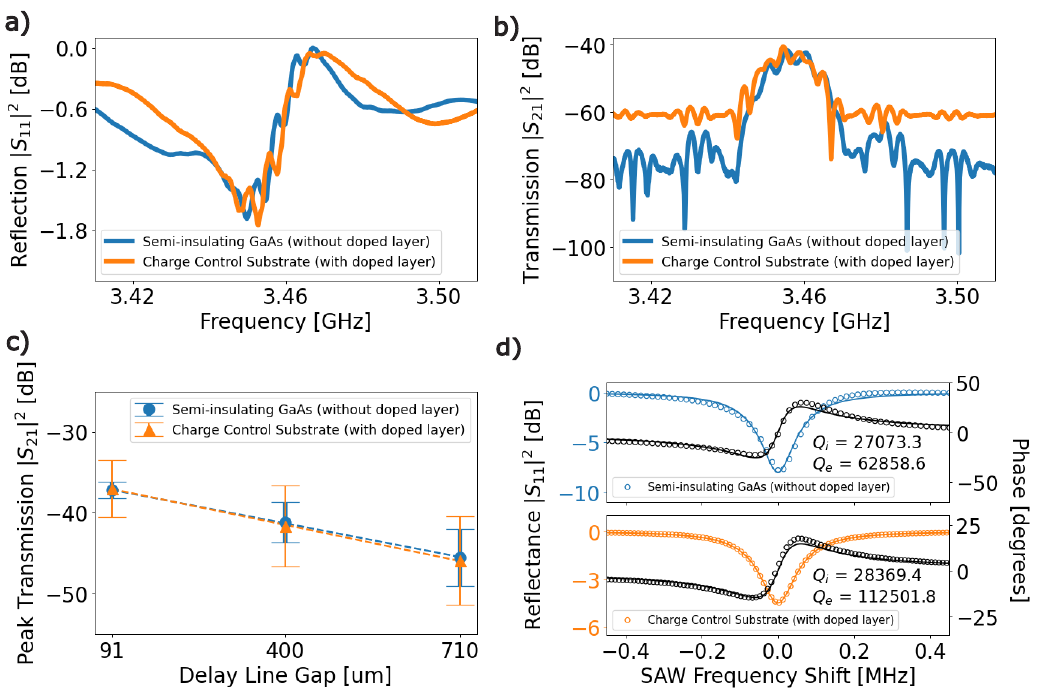}
\caption{Microwave characterization of the SAW delay lines and resonators. a) Reflection and b) transmission spectra of SAW delay lines with a 400 $\mu \text{m}$ gap. The similar peak transmission on substrates with buried n-doped GaAs compared to the same device on bulk GaAs shows the loss introduced by SAW launching and propagation is negligible with the proper choice of the depth and doping concentration of the n-doped GaAs layer. c) Peak transmitted SAW signal on the two substrates, varying the gap between the IDTs of the delay lines. The similar slope proves that the propagation loss introduced by the n-doped GaAs layer can be neglected. The error bars represent the noise in the transmission spectra as a result of the beating between the acoustic signal and electrical crosstalk. d) Reflection spectra of SAW resonators with the same design parameters, fabricated on substrates with (bottom panel) and without (top panel) the buried doped layer, showing similar internal quality factors $Q_i $$\approx$ 28,000. Circle: data. Solid line: fit according to the model described in the Supplementary Information.}
\label{fig2}
\end{figure}

\subsection*{Microwave Measurements Characterizing the SAW Resonator}

To experimentally verify these proposed solutions, we first fabricated acoustic delay lines formed by two IDTs separated by a gap of approximately 400 $\mu$m on a “charge-control substrate” (i.e., with the buried n-doped layer). We compare the delay lines’ performance to identical devices fabricated on mechanical-grade semi-insulating GaAs substrates without the n-doped layer. The microwave reflection and transmission of the acoustic delay lines are measured using a vector network analyzer (VNA) in a probe station at room temperature. As shown in Fig \ref{fig2}a), similarity in the reflection spectrum ($|S_{11}|^2$)  dip sizes between identical devices on the two substrates indicates negligible reduction in piezoelectric coupling between IDTs and SAWs. SAW generation and attenuation are analyzed by looking at peak transmitted signals in the transmission spectrum ($|S_{21}|^2$) (Fig 2b) with different delay line gaps (Fig 2c). Variations in the peak transmitted power with propagation length directly reveal SAW propagation loss rates. The error bars in Fig 2c) represent the noise in the transmission spectra as a result of the beating between the acoustic signal and electrical crosstalk. Due to the buried n-doped layer, the electrical crosstalk is larger in the “charge control substrate”, leading to slightly higher noise level. However, similar transmission peak power and loss per unit length show that (1) microwave power is delivered equally well to the IDTs in both systems and (2) propagation loss introduced by the doped layer is not evident above our measurement uncertainty. To further verify this, we fabricated and characterized losses and external coupling rates in planar SAW resonators on charge-control substrates with devices cooled to 1.7 K. By fitting the microwave reflection spectra (Supplementary Information), we extract an internal (external) quality factor $Q_i$ ($Q_e$) of approximately 28,000 (112,500) (Fig. 2d, lower panel). Both internal losses ($Q_i$) and electromechanical coupling ($Q_e$) are comparable to similar devices fabricated on a semi-insulating GaAs substrate (Fig. 2d, upper panel). In addition, the internal quality factor $Q_i$ is comparable with the state-of-the-art SAW cavities fabricated on a variety of materials at GHz frequencies \cite{Manenti2016,Moores2018,Rodriguez-Madrid2012,Shao2019,Magnusson2015,Luschmann2023}. This proves that the n-doped GaAs layer can be integrated with the SAW cavity without degrading electromechanical performance. By comparing these results with previous results in the literature \cite{DeCrescent2022}, it becomes clear that losses are dominated by features unrelated to the conductive layer, but rather to bulk scattering from the mirrors and IDTs. In particular, increased $Q_i$s and better coupling here, compared to Ref. \cite{DeCrescent2022}, are related to the lighter IDT metal (Al compared to Nb or Au; Supplementary Information).

\begin{figure}[b!]
\centering\includegraphics[width=0.9\linewidth]{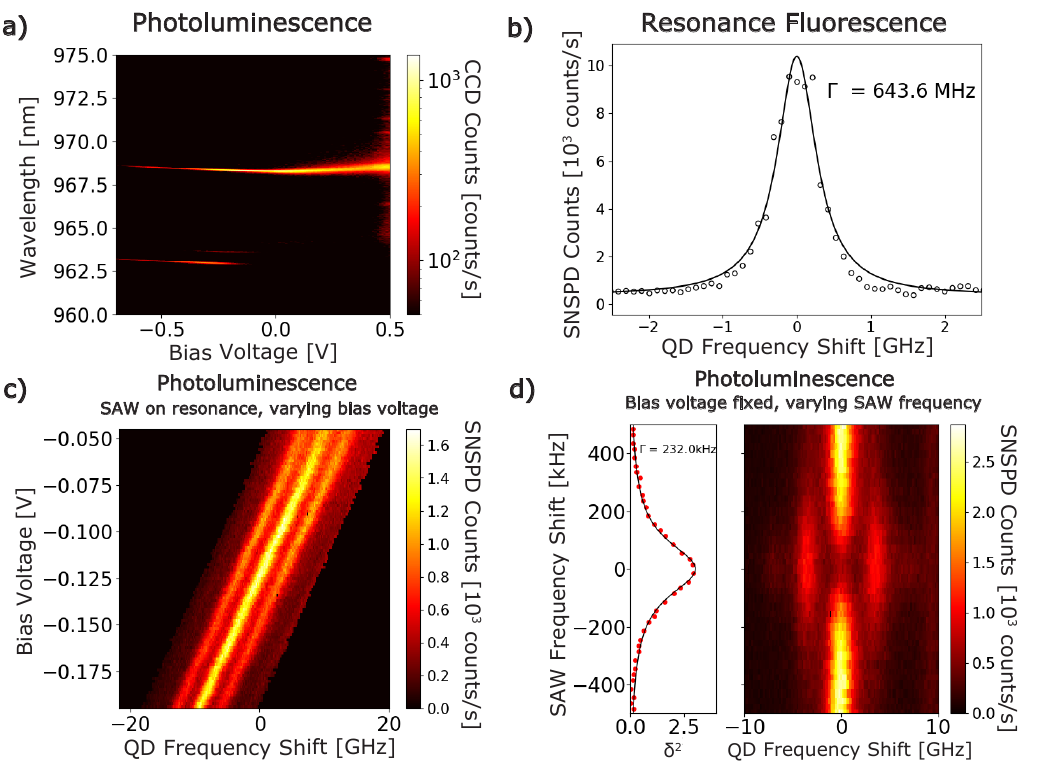}
\caption{Optical characterization of the hybrid device. a) PL spectrum of the gated QD, measured with spectrometer and CCD camera. Two charge plateaus are visible when varying bias voltage applied to the gate. b) Resonance fluorescence signal of the gated QD, measured with a superconducting nanowire single photon detector (SNSPD). A near-transform-limited linewidth is observed, which is a result of reduced charge noise by the gate structure. c) PL spectrum of the gated QD with SAW modulation at a fixed frequency (3.53388 GHz), measured with a Fabry-Perot filter and SNSPD. Both the QD’s resonant peak and sideband peaks are shifted by the Stark effect, with a rate of approximately 0.13 GHz/mV. d) PL spectrum of the gated QD at fixed gate bias with varying microwave drive frequency and constant microwave power, measured with a Fabry-Perot filter and SNSPD. A Lorentzian fit to the square of the modulation index ($\delta^2$) shows an acoustic resonance linewidth of approximately 232 kHz, corresponding to a Q factor approximately 15,000.}
\label{fig3}
\end{figure}

\subsection*{Optical Characterization of the Hybrid Device}

After addressing each challenge individually, we proceed with fabrication of the proposed hybrid device. Fig 1a) shows an optical microscope image of the fully integrated device. The complete system contains a top gate of a thin semi-transparent layer of Ti/Au inside the acoustic resonator, forming a Schottky diode structure. Optical measurements are performed on QDs directly below this thin metal gate. The thin metal layer is connected to a thick Ti/Au layer outside the acoustic resonator for wirebonding (top right in Fig 1a). The metal surrounding the SAW resonator provides ohmic contact to the n-doped layer and ensures a uniform electric potential across the layer.

The fabricated device is cooled in a 5 K optical cryostat. We perform optical and microwave measurements to characterize the optomechanical performance of the system. The behavior of the gated QDs is shown by measuring single QD photoluminescence (PL) while varying the bias voltage applied to the Schottky diode. We excite the QD with a non-resonant pump (632 nm) and measure PL with a spectrometer and a CCD detector. Fig 3a) shows two QD charge plateaus when sweeping the applied bias voltage, which is a signature of gated QDs: the emission energy changes abruptly as a result of adding or removing charges. The following experiments are conducted on the state around 0 V.

Resonance fluorescence is a very sensitive probe of the QD’s linewidth and additional noise sources in the QD’s environment \cite{Chen2016,Tomm2021}. We resonantly pump the QDs and reject reflected pump photons with a cross-polarization scheme, while fixing the bias voltage at one of the charge plateaus. The experimental setup is similar to that of Ref \cite{Imany2022}. An exemplary resonance fluorescence spectrum of the unmodulated QD is shown in Fig 3b). The measured optical linewidth is approximately 643.6 MHz, comparable to the state-of-the-art performance \cite{Tomm2021}. This narrow linewidth is a result of the reduced charge-noise of the gated QDs, compared to typical linewidth of approximately 1.7 GHz without a gate structure.

The gated QDs also interact with the SAW resonators optomechanically. We resonantly drive the SAW cavity with an external microwave source, and measure the PL of the QD as a function of bias voltage. The PL spectrum is measured with a tunable Fabry-Perot filter (600 MHz linewidth). As shown in Fig 3c), the QD is phase-modulated by the SAW field, resulting in a series of optical sidebands. The entire phase-modulated spectrum is Stark shifted by varying the bias voltage, with a rate approximately 0.13 GHz/mV, verifying that both QD charge control and SAW cavity operate simultaneously.         

The modulated QD spectrum depends on the SAW field amplitude and can act as a probe to measure the acoustic $Q$ factor of the fully integrated device under optical pumping \cite{DeCrescent2022}. The acoustic Q factor sampled by the QD includes both internal and external loss channels. We measure the QD PL spectrum with a fixed bias voltage and fixed microwave drive power while sweeping the microwave drive frequency around the SAW cavity mode. We then extract the modulation index, $\delta$, by fitting the phase-modulated PL spectrum as a function of the microwave drive frequency. A Lorentzian fit to $\delta^2$ shows a clear acoustic resonance with a linewidth of approximately 232 kHz, corresponding to a Q factor of approximately 15,000. Microwave reflection measurement of the same cavity mode shows a total $Q$ of approximately 17,800 ($Q_i \approx$ 18,500; Supplementary Information). This small discrepancy may be due to \emph{local} heating or additional charges in the cavity introduced by the pump laser, which can be avoided when pumping the QDs resonantly. This suboptimal internal quality factor of 18,500, compared to 28,000 shown in Fig 2d), is likely due to design imperfections in our cavities when introducing the thin Ti/Au layer. These design issues can be solved by more carefully compensating for SAW phase-velocity changes under the Ti/Au layer. In addition, we estimate the single-phonon coupling rate $g_0$ to be approximately 42 kHz for this device (Supplementary Information). Compared to the best $g_0$ achieved in previous work \cite{DeCrescent2022}, this smaller $g_0$ is a result of larger cavity mode volume, which was used for easier characterization. The design principle described in this work can be immediately applied to the tightly focusing cavities studied in Ref. \cite{DeCrescent2022} to achieve $g_0$ larger than 1 MHz. Therefore, the integration of gated QDs and SAW resonators can be done without sacrificing the performance of individual components. 

\section*{Discussion}

In conclusion, we have successfully incorporated low-noise gated quantum dots into high-quality-factor surface acoustic wave resonators to create fully integrated optomechanical devices, aimed at microwave-to-optical transduction. We showed that the conductive layers required for QD gating negligibly impact the SAW resonator performance. Specifically, with suitable choice of the location and the doping concentration of the doped layer, SAW generation and attenuation are virtually unaffected compared to reference systems without the QD gate layers. As a result, acoustic resonators retain excellent quality factors. Further, the additional fabrication processes required for the SAW cavities do not negatively impact the QD performance; QDs retain near transform-limited linewidths. This is critically important when operating the system as a microwave-to-optical transducer which requires illuminating the QD with a strong red-detuned pump; a narrow QD linewidth substantially reduces the rate at which this red-detuned pump directly excites the QD without phonon involvement. Therefore, the near transform-limited QDs improve the rate of getting quantum information converted to optical photons from acoustic phonons, compared to directly exciting the QD. This is a key step toward truly leveraging the excellent properties of semiconductor QDs for optomechanics.

\bibliography{reference}

\section*{Funding}
National Research Council; National Institute of Standards and
Technology.

\section*{Acknowledgments}
The authors thank Adam McCaughan for providing the Python layout design code PHIDL. The authors also thank John Teufel for the discussion on microwave transmission lines.

\section*{Data availability}
Data underlying the results presented in this paper are not publicly available at this time but may be obtained from the authors upon reasonable request.

\newpage
\section*{Supplementary Information}

\section{Sample Structure and Fabrication Process}

\begin{figure}[b!]
\centering\includegraphics[width=1\linewidth]{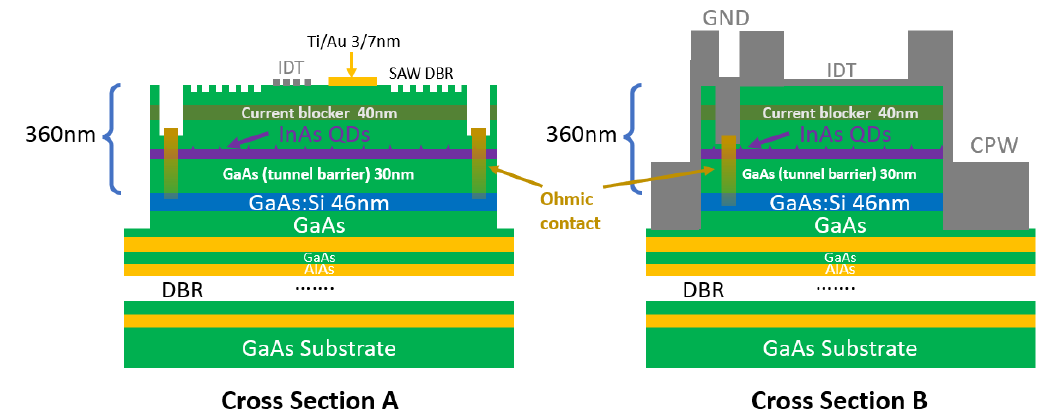}
\caption{Detailed schematic cross-sections of our devices. Cross-sections A and B correspond to the horizontal and vertical dashed lines in Fig. 1a of the manuscript.}
\label{FigS1}
\end{figure}

The layer structure of the substrate is shown in Fig S1, with cross sections A and B corresponding to the horizontal and vertical dashed lines shown in Fig 1a in the main text. All components of the vertical structure (except for the top metal layers) are grown by molecular beam epitaxy (MBE). It consists of a distributed Bragg reflector (DBR) below, formed by 22 pairs of AlAs/GaAs layers, for improved photon collection efficiency. The n-doped GaAs (GaAs:Si)  layer is 46 nm thick and located 360 nm below the surface. The InAs quantum dots (QDs) are located 30nm above the n-doed GaAs, at the antinode of the field in the weak cavity formed by the DBR and the thin Au film on top. A 40 nm layer of AlAs/GaAs superlattice is located near the surface to block current flow between the two gates of the Schottky diode (“Current-blocker”).

The substrate used in this work is grown by molecular beam epitaxy (MBE), with layer structures described in supplemental information. The MBE-grown wafer is then processed for the combined gated QD and SAW device. First, the SAW DBR (mirror) trenches are defined by electron beam lithography (EBL) and etched by reactive-ion etching (RIE). Then, ohmic contacts to the n-doped GaAs layer are made by etching vias around the SAW resonators down to approximately 100 nm above the GaAs:Si layer, and then depositing approximately 400 nm of Ni/AuGe/Ni/Au by electron-beam evaporation. The metals are thermally annealed in forming gas (95\% Ar and 5\% H2) at 420 C for 1 minute for optimum contact resistance. To finish the Schottky diode structure, a thin layer of Ti/Au (3 nm/7 nm) is evaporated inside the acoustic resonator region, along with a thick layer of Ti/Au (10 nm/250 nm) outside the resonator region to connect to large wirebond pads. In the next step, the n-doped layer is etched away everywhere outside the SAW resonator region so that the on-chip microwave coplanar waveguides (CPWs) and wirebond pads are unaffected by the doped layer. The CPW traces are then defined by photo-lithography and lifted-off with 450nm of Ti/Au. The thickness of the CPW metal is chosen to be thicker than the etched depth so that it is connected across the step. Finally, 20 nm of either Au or Al are deposited to fabricate the IDTs. The IDT structure is defined by EBL and then formed using a lift off process. We used Au IDTs for measurements shown in Fig. 2a-c and Al IDTs for measurements shown in all other figures.

\section{Theoretical estimates for SAW propagation losses and electromechanical coupling in our system.}

We use a highly n-doped thin epitaxial layer below the GaAs surface as the bottom gate of our Schottky diode structure which is used to control the electrostatic environment of our quantum dots. This conductive layer is expected to affect our device performance in two ways: 1) it may change the propagation length of a SAW by introducing new dielectric losses which interact with the electrical component of the piezomechanical wave; 2) it creates new electrostatic boundary conditions that may affect the ability of the interdigital transducer to generate piezomechanical waves in the substrate. We evaluate these two effects theoretically by performing numerical calculations using commercially available multiphysics software. Fig. S2 summarizes these calculations. In all calculations, SAWs propagate along the [110] crystal direction on the (001) surface of GaAs, and the conductive layer is 40 nm thick with uniform conductivity. 

\begin{figure}[b!]
\centering\includegraphics[width=1\linewidth]{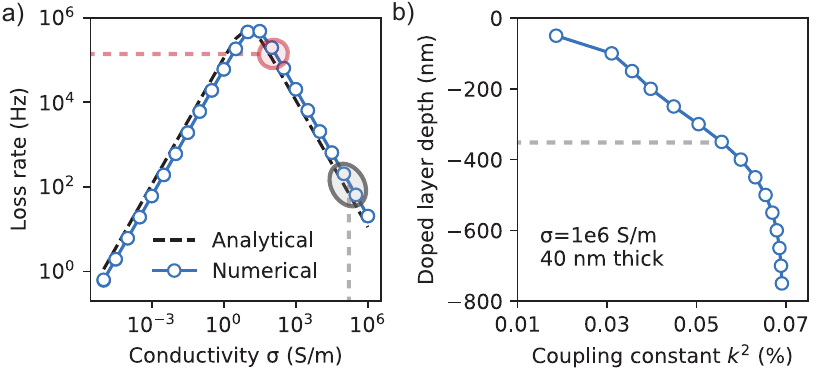}
\caption{Calculated loss rates (a) and piezoelectric coupling constants $k^2$ (b) for SAWs propagating along the [110] of a free GaAs surface above a uniform conductive thin layer (40 nm thick) with conductivity $\sigma$. a) The layer is held at 360 nm below the surface, corresponding to our fabricated sample structure, while the conductivity $\sigma$ is varied. Blue markers: numerical calculations. Black dashed line: analytical calculation. Gray ellipse: approximate loss rate expected from room-temperature conductivity measurements. Red ellipse: approximate total loss rates measured from typical SAW cavities at 3.5 GHz. b) The conductivity is held at $\sigma=10^6$ S/m while the layer depth (vertical axis) is varied. Gray dashed line: depth of the doped layer in our fabricated devices (360 nm below the surface).}
\label{figSI2}
\end{figure}

Fig. S2a shows calculated SAW propagation loss rates (blue circles) as a function of the thin layer’s conductivity $\sigma$ (constant layer depth of 360 nm). These values were derived from the complex eigenfrequencies for SAW modes calculated in a 2-dimensional unit cell with periodic boundary conditions; the loss rate is identified as the imaginary part of the complex eigenfrequency. Two distinct regimes are apparent. For $\sigma\lesssim 1$ S/m, the thin layer acts as a good (low-loss) dielectric. For $\sigma\gtrsim 10^3$ S/m, the layer acts as a good metal. Propagation loss reaches a maximum between these two regimes. These numerical results are in excellent agreement with analytical results calculated from ref. \cite{Wixforth1986} (black dashed curve). At 1.7K, our n-doped layers have characteristic conductivity of $\sigma\sim 10^5$ S/m, measured with typical Hall bar structure. These conductivities are expected to contribute approximately 100 Hz to SAW propagation losses (indicated by a gray ellipse in Fig. S2a). For comparison, SAW cavities fabricated on insulating substrates already exhibit total loss rates of approximately 100 kHz (indicated by a red ellipse in Fig. S2a). The measured losses are dominated by diffraction, impurity scattering, and bulk scattering from mirrors, other structures and possible surface roughness in our devices. The doped layer is thus expected to negligibly contribute to SAW losses in our devices. 

Fig. S2b shows calculated piezoelectric coupling constants $k^2$ as a function of the conductive layer’s depth below the surface (constant conductivity $\sigma=10^6$ S/m). The $k^2$ values were calculated by computing $k^2=2\Delta v/v$ where $\Delta v$ is the difference in SAW phase velocity between systems with electrically shorted and insulating surface boundary conditions, according to ref. \cite{aref2016}. When the depth is less than approximately 100 nm, the conductive layer effectively screens the IDT’s driving potential and strongly reduces its ability to generate SAWs. The coupling constant approaches the well-established value for insulating GaAs ($k^2\simeq0.07\%$) for depths greater than approximately 500 nm \cite{aref2016}. In our devices (360 nm, indicated by the gray dashed line), the effective coupling constant $k^2\simeq0.055\%$, approximately 80\% its value for insulating GaAs.

\section{Microwave Reflection Model}
The microwave reflection and transmission measurements shown in Fig. 2 are conducted with commercial vector network analyzer. The internal and external quality factor can then be extracted by fitting the reflection spectra according to Eqn. 1 in Ref. \cite{Manenti2016}:
\begin{align*}
    S_{11}(f) = \frac{(Q_e-Q_i)/Q_e+2iQ_i(f-f_0)/f}{(Q_e+Q_i)/Q_e+2iQ_i(f-f_0)/f}
\end{align*}
The size of the dip in $|S_{11}|^2$ reflection spectrum provides an upper bound on the electromechanical coupling efficiency, which can be improved by impedance matching with tunable superconducting microwave cavity \cite{Xu2019}.

\section{Optical Measurement Setup}
The optical measurements shown in Fig. 3 are conducted with conventional confocal microscope setup, with details described in Ref. \cite{Imany2022}. The microwave power used for measurements in Fig. 3c) and 3d) are -29 dBm and -26 dBm, respectively. For the cavity mode studied in Fig 3d), microwave reflection measurement shows a $Q_i$ of 18,396 and $Q_e$ of 548,077, corresponding to a total Q of approximately 17,798. With methods described in Ref. \cite{DeCrescent2022} appendix D, We estimate the single-phonon coupling rate $g_0$ to be approximately 42 kHz.



\end{document}